\documentstyle[sprocl,epsf]{article}

\newcommand{\nn}{\nonumber}

\def\NPA{{\em Nucl. Phys.} {\bf A}}
\def\NPB{{\em Nucl. Phys.} {\bf B}}
\def\PLB{{\em Phys. Lett.} {\bf B}}

\def\PRD{{\em Phys. Rev.} {\bf D}}
\def\PRC{{\em Phys. Rev.} {\bf C}}

\def\be{\begin{equation}}
\def\ee{\end{equation}}
\def\bea{\begin{eqnarray}}
\def\eea{\end{eqnarray}}

\def\twoder{{\stackrel{\leftrightarrow}{\nabla}}}

\begin{document}

\title{Nucleon-Nucleon Effective Field Theory at NNLO:
Radiation Pions and $^1S_0$ Phase Shift\footnote{CALT-68-2226}}

\author{Thomas Mehen and Iain W. Stewart}

\address{California Institute of Technology, Pasadena, CA, 91125\\
E-mail: mehen@theory.caltech.edu, iain@theory.caltech.edu}

\maketitle\abstracts{Low energy phenomena involving two nucleons can be
successfully described using effective field theory.  Because of the relatively large
expansion parameter, it is only at next-to-next-to-leading order (NNLO) where
one can expect to see agreement with experiment at the few percent level.  The
first part of this talk will focus on radiation pion effects, which first appear at
NNLO. The power counting for radiation pions is simple for center of mass
momentum $p \sim \sqrt{M m_\pi}\equiv Q_r$, the threshold for pion production. 
We explain how graphs calculated with the $Q_r$ power counting scale for $p\sim
m_\pi$.  The $Q_r^3$ radiation pion contributions to nucleon-nucleon scattering
are suppressed by inverse powers of the S-wave scattering lengths.  However, we
point out that order $Q_r^4$ radiation contributions might give a NNLO
contribution for $p \sim m_\pi$.  In the second part of the talk, results for the
potential pion and contact interaction part of the NNLO  $^1S_0$ phase shift are
presented.  We emphasize the importance of eliminating spurious poles in the
expression for the amplitude at each order in the perturbative expansion. Doing
this leaves a total of three free parameters at NNLO.  We obtain a good fit to the
$^1S_0$ phase shift.}

\section*{Introduction}

This talk focuses on higher order calculations in the low energy effective field
theory for two-nucleon systems. In particular, we will be discussing
nucleon-nucleon scattering at next-to-next-to-leading order (NNLO) in the
expansion recently proposed by Kaplan, Savage and Wise (KSW).\cite{ksw1,ksw2}
Observables are expanded in powers of $Q/\Lambda$, where $Q$ is either $p$, the
three-momentum of the two nucleons in the center of mass frame, or $m_\pi$. 
$\Lambda$ is the range of the effective theory. The nuclear S-wave scattering
lengths (denoted by $a$) are very large so that powers of $p a$ have to be
summed to all orders at each order of the $Q$ expansion. This requires a novel
power counting in which the leading 4-nucleon operator with no derivatives is
treated nonperturbatively.  To make this power counting manifest in dimensional
regularization it is necessary to use subtraction schemes such as
PDS\,\cite{ksw1,ksw2} or OS\,\cite{ms0,ms1}, but predictions of the theory are
manifestly scheme and scale independent order by order in $Q$.  Higher derivative
operators and pion exchange are treated perturbatively.  

Various estimates of the range of the theory exist. One estimate comes from
examining pion exchange ladder graphs, where each additional loop gives a
contribution of order $p \times M g_A^2/(8 \pi f^2) \equiv p/\Lambda_{NN}$, where
$\Lambda_{NN} = 300 \,\rm{MeV}$. Another possibility is that $m_\rho$ or the
threshold for $\Delta$ production sets the scale for the breakdown of the effective
theory, implying a range for S-wave scattering $\sim 700 \,\rm{MeV}$.  It has been
suggested \cite{lepage,steele} that two pion exchange contributions to the
nucleon-nucleon potential may be become important for momenta of order $400
\,\rm{MeV}$. These considerations point to a range somewhere between
$300\,\rm{MeV}$ and $700 \,\rm{MeV}$.  Therefore, for $p\sim m_\pi$, the expansion
parameter, $Q/\Lambda$, is between $1/2$ and $1/5$.  Because the expansion
parameter of the theory is rather large, low order calculations in the effective
theory cannot be expected to reproduce phase shift data as accurately as potential
models with many parameters.  

Many observables have been computed to NLO in the KSW expansion, including
nucleon-nucleon phase shifts,\cite{ksw1,isospin} Coulomb corrections to
proton-proton scattering,\cite{ppsca} electromagnetic form factors of the
deuteron,\cite{demf} deuteron polariz\-abilities,\cite{dpol} proton-proton
fusion,\cite{ppfus} $n p\rightarrow d \gamma$,\cite{npdg} Compton deuteron
scattering,\cite{dC} and $ \nu d \to \nu d$.\cite{butler} Some of these calculations
are reviewed in the talk by Martin Savage in this volume.\cite{NLO} One typically
finds errors of order $30-40 \%$ at leading order and $10 \%$ at NLO.  This is
consistent with $Q / \Lambda \approx 1/3$, or $\Lambda \approx 400\,{\rm MeV}$.
This suggests that at NNLO, effective field theory calculations of low energy
processes in the two body sector should agree with data at the few percent level,
approaching an accuracy comparable to that of potential models. It is for this
reason that extending calculations to this order is an important part of the effective
field theory program.  

In the first half of this talk, we will discuss radiation pion effects, which first appear
at NNLO in calculations of nucleon-nucleon scattering. The power counting of
KSW has to be modified in the presence of pion radiation because a new scale,
$Q_r = \sqrt{M m_\pi}$, appears.\cite{radpi} For power counting radiation pions, it is
simpler to take $p \sim Q_r$ and count powers of $Q_r$ rather than $Q$. We give
a procedure for determining how a $Q_r^n$ correction scales with $Q$ for $p\sim
m_\pi$.  The order $Q_r^3$ radiation pion contribution to nucleon-nucleon
scattering turns out to be suppressed by powers of $1/a$.  This is actually a
consequence of the invariance of the leading order theory under Wigner's $SU(4)$
spin-isospin symmetry.\cite{msw} Wigner symmetry is discussed in detail in the
talk by Mark Wise in this volume.\cite{Wigner}  The order $Q_r^4$ radiation pion
corrections can give an order $Q$ contribution. In the second half of the talk, we
present results of a NNLO calculation of nucleon-nucleon scattering in the
$^1S_0$ channel.  It is emphasized that coupling constants of the theory must be
treated in a $Q$ expansion. The parameter space is constrained by the requirement
that perturbative corrections do not shift the location of the pole in the amplitude
and by the solutions of renormalization group equations.  Once these constraints
are imposed, a three parameter fit to the $^1S_0$ phase shift is demonstrated
which has $<2\%$ accuracy at $p\sim m_\pi$ and also reproduces the data well for
higher momenta. A similar calculation is discussed in the talk by Gautum
Rupak\,\cite{NNLO,NNLO2}.  

\section*{Radiation and Soft Pions}

The Lagrangian for the theory of nucleons and pions is 
\begin{eqnarray}\label{Lag}
{\cal L}_\pi &=& \frac{f^2}{8} {\rm Tr}\,( \partial^\mu\Sigma\: \partial_\mu
\Sigma^\dagger )+\frac{f^2\omega}{4}\, {\rm Tr} (m_q \Sigma+m_q 
\Sigma^\dagger)+ \frac{ig_A}2\, N^\dagger \sigma_i (\xi\partial_i\xi^\dagger -
\xi^\dagger\partial_i\xi) N  \nn\\ &+& N^\dagger \bigg( i
D_0+\frac{\vec D^2}{2M} \bigg) N - {C_0^{s}} ( N^T P^{s}_i
N)^\dagger ( N^T P^{s}_i N) \label{Lpi} \\ &+&   {C_2^{s}\over
8} \left[ ( N^T P^{s}_i N)^\dagger ( N^T P^{s}_i
\:\twoder^{\,2} N) + h.c. \right] \nn\\ 
&-&{D_2^{s}}\, \omega\, {\rm Tr}(m^\xi ) ( N^T P^{s}_i N)^\dagger ( N^T P^{s}_i N) 
+ \ldots \nn \,,
\end{eqnarray}
where operators relevant at NLO are included (and isospin violation is neglected). 
Here $g_A=1.25$ is the nucleon axial-vector coupling, $\Sigma = \xi^2$ is the
exponential of pion fields, $f=131\, {\rm MeV}$ is the pion decay constant,
$m^\xi=(\xi m_q \xi + \xi^\dagger m_q \xi^\dagger)/2$, where $m_q={\rm
diag}(m_u,m_d)$ is the quark mass matrix, and $m_\pi^2 = w(m_u+m_d)$.  The
matrices $P_i^{s}$ project onto states of definite spin and isospin, and the
superscript $s$ denotes the partial wave amplitude mediated by the operator.  This
talk will be concerned only with S-wave scattering, so $s=S$ (for ${}^1S_0$) or
$T$ (for ${}^3S_1$).  This notation will be omitted when it is not necessary to
distinguish between the two channels.

In the KSW power counting, the $C_0$ operator is treated nonperturbatively and
graphs with a single pion exchange (dressed with $C_0$ bubbles) first appear
at NLO.  Loop graphs with pions contain three different kinds of contributions,
which are called potential, radiation and soft. The three kinds of pion are
characterized by different energy $(q_0)$ and momentum $({\vec q\,})$:
\begin{eqnarray} 
&{\rm potential} \qquad\quad &q_0 \sim {\vec q\,}^2/M \nn \\
&{\rm radiation} \qquad\quad &q_0 \sim |{\vec q\,}| \sim m_\pi \nn \\ 
&{\rm soft} \qquad\quad &q_0 \sim |{\vec q\,}| \sim Q_r =\sqrt{M m_\pi} .\nn 
\end{eqnarray} 
As stated earlier, when calculating radiation and soft contributions we take $p \sim
Q_r$ rather than $p\sim Q$.  The three contributions will differ in size and it is
necessary to devise a power counting which correctly takes this into account. 
Before giving the power counting we will illustrate how these contributions arise
with a few illustrative examples.

\begin{figure}[t!]
  \centerline{\epsfysize=3.0truecm \epsfbox{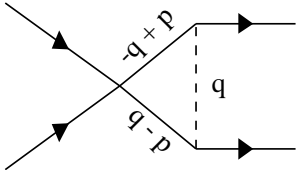}  }
 {\caption[1]{One loop graph with a $C_0$ and pion that has both potential and 
radiation contributions} 
  \label{fig1} }
\end{figure}

Consider the one loop graph shown in Fig.~\ref{fig1}, 
whose contribution to the amplitude is proportional to 
\begin{eqnarray}\label{oneloop} 
  {g_A^2 C_0 \over 2 f^2} \int\!\! {d^4 q \over (2\pi)^4 }
   {1 \over \frac{E}2+q_0 -{({\vec q\,}-{\vec p})^2 \over 2M} +i
  \epsilon}\: {1 \over \frac{E}2-q_0 -{({\vec q\,}-{\vec p})^2 \over 2M} +i
  \epsilon}\: {{\vec q\,}^2 \over q_0^2 - {\vec q\,}^2 - m_\pi^2 + i \epsilon} 
  \nonumber  
\end{eqnarray} 
When the $q_0$ integral is performed via contour integration, the integral receives
contributions from both pion and nucleon poles. If a nucleon pole is taken then
$|q_0| = E/2 - ({\vec q\,}-{\vec p\,})^2/2M  =  (2{\vec q\,}\cdot{\vec p\,}-{\vec
q\,}^2)/2M$, where in the last step we have used $E = {\vec p\,}^2/M$.  Since $|{\vec
p}|,|{\vec q\,}| \ll M$, $q_0 \ll |{\vec q\,}|$, we can expand the pion propagator:
\begin{eqnarray}
 {1 \over q_0^2 - {\vec q\,}^2 - m_\pi^2}= -{1 \over {\vec q\,}^2 + m_\pi^2}- 
  {q_0^2 \over ({\vec q\,}^2 + m_\pi^2)^2}+... \,. \nonumber
\end{eqnarray} 
A pion is referred to as a potential pion whenever the energy dependent piece of its
propagator is treated perturbatively.  In the KSW power counting, $|{\vec p\,}| \sim
m_\pi \sim \mu_R\sim |{\vec q\,}| \sim Q$. Each nucleon propagator gives a factor of
$M/Q^2$ since $E \sim q_0 \sim Q^2/M$.  The measure $d^4 q \sim Q^5/M$, and in
a scheme with manifest power counting, $C_0 \sim 1/(M\mu_R) \sim 1/(MQ)$. Using
this counting it is straightforward to see that this graph is order $Q^0$, i.e., it is a
NLO contribution. The first correction from the expansion in $q_0^2/(\vec{q\,}^2
+m_\pi^2)$ is suppressed relative to the leading potential contribution by
$Q^2/M^2$, and so is N$^3$LO.  

There is also a contribution from the pion pole. In this case $q_0^2 = {\vec q\,}^2
+m_\pi^2$. In the nucleon propagators, the factors of $(2{\vec q\,}\cdot{\vec
p\,}-{\vec q\,}^2)/2M \ll q_0$ and must be treated perturbatively.  With the KSW
power counting, the nucleon propagators in the graph in Fig.\ 1 are $\pm 1/q_0 \sim
1/Q$, and the loop measure scales as $Q^4$, so the graph is $\sim Q$. 
Therefore, this radiation pion contribution first appears at NNLO.

\begin{figure}[t!]
  \centerline{ \epsfysize=4.0truecm \epsfbox{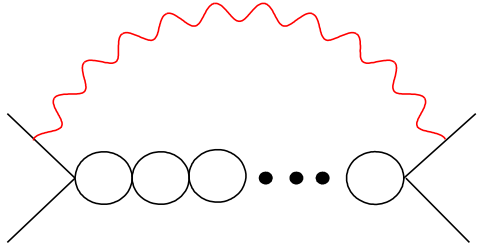}  }
 {\caption[1]{A radiation pion graph with $n$ internal $C_0$ bubbles.} 
  \label{fig2} }
\end{figure}

While KSW power counting works for the graph in Fig.~\ref{fig1}, it fails for 
other graphs with radiation pions.  As an example, consider the graph in 
Fig.~\ref{fig2}
which contains $n$ nucleon bubbles inside a radiation pion loop.  The loop integral
in this graph vanishes if the pion pole is not taken so there is no potential pion
contribution.  Emission of the radiation pion in these graphs changes the
spin/isospin of the nucleon pair.  Therefore, if the external nucleons are in a
spin-triplet (singlet) state, then the coefficients appearing in the internal bubble
sum are $C_0^{S}\:(C_0^{T})$. For definiteness, consider nucleon-nucleon
scattering in the $^1S_0$ channel.  The contribution from the graph in 
Fig.~\ref{fig2} is:
\begin{eqnarray}\label{gr3a}
&& {g_A^2 \over 2 f^2} \int\!\! {d^4 q \over (2 \pi)^4} {i \over q_0 + i \epsilon}\: {i
\over q_0 + i \epsilon}\: {-i \,{\vec q\,}^2 \over q_0^2 -{\vec q\,}^2 - m_\pi^2 +
i\epsilon} [- i C_0^T(\mu_R)]^{n+1}  \nn \\
 &&\qquad\qquad \times \left[ \int\!\! {d^4 k \over (2 \pi)^4} {i \over q_0 -k_0 +
  \frac{E}2-{({\vec k}-{\vec q\,})^2  \over 2M}+i \epsilon}\: {i \over k_0 +\frac{E}2-
  {{\vec k}^2 \over 2M} +i \epsilon} \right]^n \,. \nn
\end{eqnarray}
The $q_0$ integral is closed around the one pion pole above the real axis so $q_0
\sim |{\vec q\,}| \sim Q$. In the $k_0$ integrals a nucleon pole must be taken. In the
KSW power counting, $k_0 \sim Q^2/M, |{\vec k\,}| \sim Q$. The graph then scales
as $(Q/M)^{n+1}$. This suggests that graphs with nucleon bubbles inside the
radiation loop are suppressed relative to the one loop radiation pion graph.
However, explicitly performing the $q_0$ and $k_0$ integrals gives
\begin{eqnarray}
   -iC_0^T(\mu_R)\,{g_A^2 \over 2 f^2}\!\! \int\!\!\! {d^3 q \over (2\pi)^3} 
  {{\vec q\,}^2 \over ({\vec q\,}^2 + m_\pi^2)^{3/2}} \left[ \int\!\! {d^3 k \over (2 \pi)^3}
  {-M C_0^T(\mu_R)  \over {\vec k\,}^2+ M({\vec q\,}^2 + m_\pi^2)^{1/2}-ME}
   \right]^{n} \nn \!\!\!\! .
\end{eqnarray} 
The size of the loop momenta $k$ in the nucleon bubbles is $\sim \sqrt{M
m_\pi}$ even for $p < \sqrt{M m_\pi}$.  The integral will be dominated by ${\vec q\,} 
\sim m_\pi$ so the graph will scale as
\begin{eqnarray}
{1\over \Lambda_\chi^2}\:{m_\pi^2 \over M\mu_R}\bigg({\sqrt{M m_\pi} \over 
\mu_R }\bigg)^n \,. \nn
\end{eqnarray}
(Recall $\Lambda_\chi = 4 \pi f$).  For $\mu_R \sim Q \sim m_\pi$, we see that
graphs with additional bubbles are enhanced, contrary to the KSW power counting.
The sum over graphs with an arbitrary number of bubbles is $\mu_R$ independent
and the correct estimate for the size is obtained when $\mu_R \sim \sqrt{M m_\pi}$.
At this scale, these graphs and their sum are of order $m_\pi^{3/2}/(M^{3/2} 
\Lambda_\chi^2)=Q_r^3/(M^3\Lambda_\chi^2)$.

In the KSW power counting, one assumes the loop momentum in the nucleon
bubbles is dominated by $k \sim Q, k_0 \sim Q^2/M$. However, when the nucleon
bubbles are inside a radiation pion loop, the energy flowing into the nucleon
bubbles is actually order $m_\pi$, and therefore $k^0\sim m_\pi$ and $k \sim
\sqrt{M m_\pi} \equiv Q_r$.  This scale corresponds to the threshold for on-shell
pion production.  

In general, radiation pion graphs will depend on $p$, $m_\pi$ and $M$ in a
complicated way, making them difficult to power count if one takes $p \sim m_\pi$.
However, the natural scale for loop momenta with radiation pions is $Q_r$, and the
power counting simplifies considerably if we consider nucleons scattering at $p
\sim Q_r$.  Later we will discuss what happens as $p$ is lowered back down to
$m_\pi$.

\begin{figure}[!t]
  \centerline{\epsfysize=6.0truecm \epsfbox{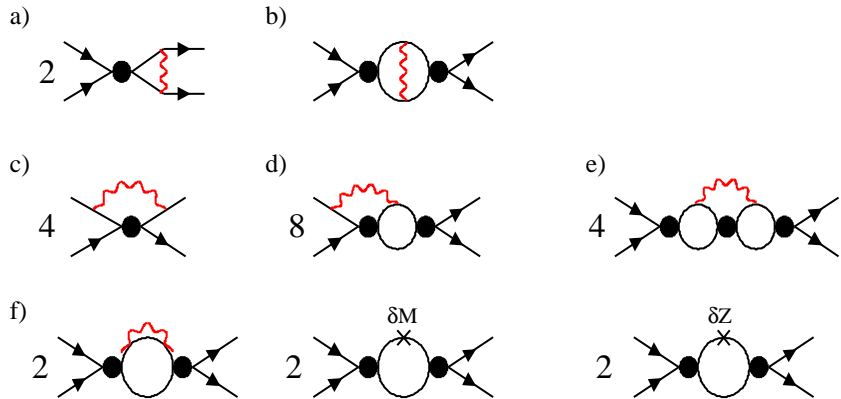}  }
 {\caption[1]{Leading order radiation pion graphs for $NN$ scattering.  The solid
lines are nucleons, the wavy lines are radiation pions and $\delta M$, $\delta Z$ 
are the mass and field renormalization counterterms. The filled dot denotes the
$C_0(\mu_R)$ bubble chain.  There is a further field renormalization contribution
that is not shown, but is included in the calculation.\cite{radpi}} \label{Qr3} }
\end{figure}

Power counting at the scale $Q_r$ is straightforward.  We take $p \sim \mu_R \sim
Q_r$. The $C_{2n}$ scale with $\mu_R$ exactly the same way as in the KSW
power counting, $C_{2n}p^{2n} \sim p^{2n}/\mu_R^{n+1} \sim Q_r^{n-1}$. A
radiation pion propagator gives $M^2/Q_r^4$, the pion nucleon coupling gives
$Q_r^2/M$. Nucleon propagators scale like $M/Q_r^2$.  In a radiation loop $q_0
\sim |{\vec q}\,| \sim m_\pi$, so the loop measure $d^4q \sim Q_r^8/M^4$. The
measure of a potential loop scales as $Q_r^5/M$. Using this power counting it is
straightforward to show that all graphs with one radiation pion and an arbitrary
number of $C_0$'s scale as $Q_r^3/(M^3 \Lambda_\chi^2)$.  These graphs are
shown in Fig.~\ref{Qr3}.

It is interesting to examine the result of evaluating some of the diagrams in 
Fig.~\ref{Qr3} explicitly\,\cite{radpi} $(\bar \mu^2 = \mu^2 \pi e^{-\gamma_E})$:
\begin{eqnarray}\label{gr1} 
a)&=& 
  -3 i {\cal A}_{-1} { g_A^2 m_\pi^2\over (4 \pi f)^2} \bigg[ {1 \over \epsilon} + 
  {1\over 3} - {\rm ln}\Big({m_\pi^2 \over {\overline{\mu}}^2}\Big) \bigg] \,,  \\
b) &=& [{\cal A}_{-1}]\,^2\, {g_A^2 M m_\pi^2 \over (4 \pi f)^2}\, \Bigg\{
   {3\,p \over 4 \pi}\, \bigg[ {1 \over \epsilon} + {7\over 3} -
   2\,{\rm ln}\, 2 - {\rm ln}\Big({m_\pi^2 \over {\overline{\mu}}^2}
   \Big) - {\rm ln}\Big({-p^2 \over {\overline{\mu}}^2}\Big) \bigg]  \nn \\
 &&\qquad\qquad\qquad\qquad +{i\sqrt{M m_\pi} \over 4\sqrt{\pi}}\,I_1 \Big({E\over m_\pi}
   \Big) \Bigg\} \,, \nn \\
c) &=& {ig_A^2 \over \sqrt{\pi} f^2}  \Big({m_\pi \over M}\Big)^{3/2} 
    I_2 \Big( {E \over m_\pi} \Big)\,. \nn
\end{eqnarray}
$I_1$ and $I_2$ are hypergeometric functions. The $1/\epsilon$ poles are cancelled
by insertions of a $D_2 m_\pi^2$ counterterm.  The leading order amplitude
\begin{eqnarray}
 {\cal A}_{-1} = -{4\pi \over M} {1 \over 1/a^S+i p} \,, \nn 
\end{eqnarray}
scales as $\sim 1/(Mp)$, so we see that Eq.~(\ref{gr1}) has terms proportional to 
\begin{eqnarray}
\left({m_\pi \over M}\right)^{3/2} \!\!\!\!\!\!, \qquad {m_\pi^2 \over M p} \qquad 
{\rm and} \qquad {m_\pi^{5/2} \over M^{1/2} p^2}\,. \nn
\end{eqnarray}
For $p \sim Q_r$ these terms scale as $Q_r^3/M^3$, as anticipated by the power
counting. At $p \sim m_\pi \sim Q$, these terms scale like $(Q/M)^{3/2}, Q/M$, and
$(Q/M)^{1/2}$ respectively.  Bubble sums which do not appear inside radiation
loops will be referred to as external bubble sums.  These bubble sums are 
responsible for the factors of $p$ in the denominators, and thus the enhancement
of some terms at low momentum.  Graphs with two external bubble sums have
terms that are enhanced by $Q_r^2/Q^2\sim 1/m_\pi$ (and $Q_r/Q \sim 
1/\sqrt{m_\pi}$), while graphs with one external bubble sum have terms enhanced
by $1/\sqrt{m_\pi}$.  Individual graphs like $b)$ have parts that scale differently
with $Q$. Terms which scale like $Q^{1/2}$ at low $p$ are actually larger than
NNLO in the $Q$ counting.  The $Q^{1/2}$ contributions come from graphs $b), e)$
and $f)$, and cancel when the graphs are added together.  Presently it is not
known whether this cancellation occurs for some reason or is merely an accident.

The sum of all $Q_r^3$ graphs is:
\begin{eqnarray} \label{fans}
 i\,{\cal A}^{rad}&=& 6i\, {\cal A}_{-1}^2 {g_A^2 m_\pi^2 \over (4 \pi f)^2}\, 
  {M  \over 4\pi}\left( {1\over a^S} -{1\over a^T}\right)\, \bigg[   \kappa + 
  {\rm ln}\Big({ {\mu}^2\over m_\pi^2 }\Big) \bigg]   \nn \\
 && +i\,  {\cal A}_{-1}^2 \left({M\over 4\pi}\right)^2 \left( {1\over a^S} -
  {1\over a^T}\right)^2  {g_A^2 \over \sqrt{\pi} f^2} \Big({m_\pi \over M}\Big)^{3/2} 
  I_2 \Big( {E \over m_\pi} \Big)\,, 
\end{eqnarray}
where the $\ln(\mu)$ dependence is cancelled by $D_2(\mu)$.  (For the $^3S_1$
channel, the result is the same as above with $a_S \leftrightarrow a_T$.) The final
answer turns out to be much smaller than anticipated by the power counting.  For
$p \sim Q_r$, the first term is suppressed by a factor of $1/Q_r(1/a^S-1/a^T)$, the
second by $1/Q_r^2(1/a^S-1/a^T)^2$. This suppression occurs because the
radiation pions couple to a charge of Wigner's $SU(4)$, which is a symmetry of the
leading order Lagrangian in the limit $a_S, a_T \rightarrow \infty$ (or $a_S =
a_T$).\cite{msw,Wigner} The order $Q_r^3$ radiation pion graphs are a tiny
correction to the S-wave scattering amplitude.  

The next important radiation pion contribution comes from graphs with one
insertion of a $C_2 p^2$, $D_2 m_\pi^2$, or $G_2$ operator \cite{ms1} or a
potential pion, and one radiation pion with an arbitrary number of $C_0$'s. Power
counting these graphs gives $Q_r^4/(M^3 \Lambda_\chi^2 \Lambda)$, i.e. these are
suppressed by $Q_r/\Lambda$ relative to the leading radiation pion graphs in Fig.\
\ref{Qr3}. Note that $Q_r = 360 \,{\rm MeV}$, so for the most pessimistic estimates
of $\Lambda$, the $Q_r/\Lambda$ expansion does not converge.   If this is the case
then the radiation pion contribution is incalculable. This is true of radiation
contributions even when we scale down to $p\sim m_\pi$. However, at the low
momenta where the theory would be applicable, the radiation pions could be
integrated out.  For example, the hypergeometric function $I_2(E/m_\pi)=I_2(p^2/(M
m_\pi))$ could be expanded in a series in $p^2$ and the effect of each term
absorbed\footnote{Unfortunately, the resulting theory below the scale $Q_r$
would no longer respect chiral symmetry.  Operators involving a different number
of pion fields could have different coefficients.} into the definition of a $C_{2 n}$. 
Assuming the radiation pion contribution is computable, the $Q_r^4$ correction is
almost certainly larger than the $Q_r^3$ correction. Since the $C_2 p^2$, $D_2
m_\pi^2$ operators and potential pion exchange do not respect Wigner's $SU(4)$,
there will be no suppression by factors of $1/(a\,Q_r)$.  

An important issue which needs to be addressed is how the radiation pions graphs
scale as $p$ is lowered from $Q_r$ to $m_\pi$. We saw that the $Q_r^3$ graphs
had pieces that scaled as $Q^{1/2}, Q$, $Q^{3/2}, \ldots$, for $p \sim m_\pi$, and
that this enhancement can be understood by counting the number of external
bubble sums. In order to know which radiation pion graphs to include at a given
order in the KSW power counting, we must know how a $Q_r^n$ correction scales
with $Q$ for $p \sim m_\pi$.  It turns out that an order $Q_r^n$ calculation is 
sufficient to determine the order $Q^{n/2-1}$ result.

To see this first consider the $Q$ expansion of $p \, {\rm cot}\, \delta$:
\begin{eqnarray}
p \, {\rm cot}\, \delta &=& i p + {4 \pi \over M} {1 \over {\cal A}} \nn \\
  &=& i p + {4 \pi \over M} {1 \over {\cal A}_{-1}} 
  - {4 \pi \over M} {{\cal A}_0 \over  {\cal A}_{-1}^2} 
  - {4 \pi \over M} \left( {{\cal A}_1 \over  {\cal A}_{-1}^2} - 
  {{\cal A}_0^2 \over  {\cal A}_{-1}^3} \right) \nn \\*
  &&- {4 \pi \over M} \left( {{\cal A}_2 \over  {\cal A}_{-1}^2} - {2 {\cal A}_0 
  {\cal A}_1 \over  {\cal A}_{-1}^3}+{{\cal A}_0^3 \over  {\cal A}_{-1}^4} \right)+
  \dots \,. \nn
\end{eqnarray}
$p\, {\rm cot}\, \delta$ is real and an analytic function of $p^2$ near $p=0$. 
This will be true order by order in $Q$ so:
\begin{eqnarray}
{{\cal A}_0 \over  {\cal A}_{-1}^2} = f_0 \, &\Rightarrow& \, 
    {\cal A}_0 = f_0  {\cal A}_{-1}^2  \,, \nn \\
{{\cal A}_1 \over  {\cal A}_{-1}^2} - {{\cal A}_0^2 \over  {\cal A}_{-1}^3} 
    = f_1 \, &\Rightarrow& \, {\cal A}_1 = f_1  {\cal A}_{-1}^2+f_0^2  {\cal A}_{-1}^3
    \,, \nn \\
{{\cal A}_2 \over  {\cal A}_{-1}^2} - {2 {\cal A}_0 {\cal A}_1 \over 
      {\cal A}_{-1}^3}+{{\cal A}_0^3 \over  {\cal A}_{-1}^4} = f_2 \, &\Rightarrow& \,
     {\cal A}_2 = f_2  {\cal A}_{-1}^2+2 f_0 f_1  {\cal A}_{-1}^3 
    + f_0^3  {\cal A}_{-1}^4 \nn \,,
\end{eqnarray}
where the $f_n$ are real functions of $p$ which are analytic about $p^2 = 0$. We
see that the general form of a higher order amplitude is powers of ${\cal A}_{-1}$
multiplied by functions of $p$.  The crucial point is that the function multiplying
the ${\cal A}_{-1}^2$ is the only new contribution. The coefficient of ${\cal
A}_{-1}^n, n > 2$, is determined by lower order amplitudes. In the $Q$ expansion 
of $p\, {\rm cot}\, \delta$ the latter contributions will cancel.

\begin{figure}[!t]
  \centerline{\epsfysize=2.3truecm \epsfbox{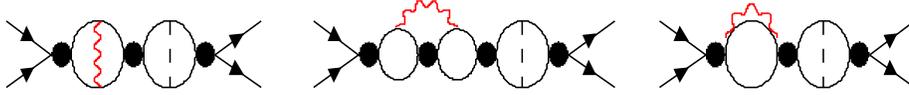}  }
 {\caption[1]{Example of order $Q_r^4$ graphs that have three external bubble
sums.} \label{Qr4a} }
\end{figure}

This generalizes to the $Q_r$ expansion of radiation pion graphs, the only
difference being that the radiation pion contribution starts out at $Q_r^3$, while
the potential pion starts out at $Q_r^0$. A $Q_r^n$ radiation pion correction to the
amplitude will be of the form:
\begin{eqnarray}
  {\cal A}_{n} = {\cal A}_{-1}^2 f_{n,2} + {\cal A}_{-1}^3 f_{n,3} + 
  \ldots + {\cal A}_{-1}^{n-1} f_{n,n-1} . \nn
\end{eqnarray}
Again, the $f_{n,m}$ are analytic about $p^2 = 0$ and all the $f_{n,m}$ except for
$f_{n,2}$ will be determined from lower order amplitudes.  Since ${\cal A}_n
\sim Q_r^n$ and ${\cal A}_{-1} \sim 1/(M p)$, $f_{n,m} \sim Q_r^{n+m}$ for $p \sim
Q_r$. To understand how $f_{n,m}$ scales with $Q$ as $p$ is lowered to $m_\pi$, 
note that without loss of generality, $f_{n,m}$ can be written as
\begin{eqnarray}
  f_{n,m} = { ({\sqrt{M m_\pi}})^{n+m} \over \Lambda_\chi^2 \ 
  {\overline \Lambda}^{\:n-m} } {\hat f}_{n,m} \left({p \over \sqrt{M m_\pi}}, 
  \ldots \right) \nn \,,
\end{eqnarray}
where the ellipses denote momentum dependence that involves scales other than 
$Q_r$, and ${\overline \Lambda}=\Lambda_\chi$, $\Lambda$, or $M$. For 
$p\sim m_\pi$ the ellipse denote dependence on the dimensionless
variables $p/m_\pi$, $p a$, and $p/{\overline \Lambda}$. For $p \sim m_\pi$, 
${p/\sqrt{M m_\pi}} \sim (Q/M)^{1/2}$ and the function ${\hat f}_{n,m}$ can be 
expanded in its first argument:
\begin{eqnarray}
  {\cal A}_{-1}^m f_{n,m} = {\cal A}_{-1}^m { ({\sqrt{M m_\pi}})^{n+m} \over 
  \Lambda_\chi^2 \ {\overline \Lambda}^{\:n-m} } {\hat f}_{n,m}
  \left( 0, \ldots \right)\left[1 + O \left( {Q \over M} \right)^{1/2} \right] \,. \nn
\end{eqnarray}
The leading term scales as $\sim Q^{n/2-m/2}$ for $p \sim m_\pi \sim Q$.  Since
only the $m=2$ term actually ends up contributing to $p\,{\rm cot} \, \delta$, the
new contribution at $Q_r^n$ scales like $Q^{n/2-1}$ (plus subleading terms) for
$p \sim m_\pi$.  This is consistent with the result of the $Q_r^3$ calculation, where
the largest contributions from individual graphs scaled as $Q^{1/2}$.  A
cancellation between graphs resulted in this contribution vanishing.  The remaining
terms scale as $Q$, $Q^{3/2}, \ldots$ (counting $1/a\sim Q$). The $Q_r^n$
radiation pion contribution could have a $Q^{1/2}$ contribution from the ${\cal
A}_{-1}^{n-1} f_{n,n-1}$ term. But this contribution is determined by the $Q_r^3$
amplitude which vanished; so there will be no $Q^{1/2}$ contribution from
any radiation pion graph. For example, at $Q_r^4$ the terms proportional to ${\cal
A}_{-1}^3$ comes from graphs such as those shown in Fig.~\ref{Qr4a}.  
These graphs factorize into two pieces which are lower order and it is easy to see
that the same cancellation between the $Q^{1/2}$ pieces of the graphs $b), e)$ and
$f)$ will also occur at $Q_r^4$.  

Since $Q^{n/2-1} = Q$ for $n=4$, the $Q_r^4$ radiation pion graphs may have a
contribution that is NNLO for $p \sim m_\pi$.  We have checked that graphs with
one insertion of the $D_2 m_\pi^2$ operator do not give rise to such a contribution,
but a calculation of the remaining $Q_r^4$ graphs has not been performed.  These
graphs need to be computed in order to obtain the complete NNLO amplitude. 
Unfortunately, graphs with one potential and one radiation pion are numerous.  
Higher $Q_r^n$ amplitudes may have a contribution which scales as $Q$ for $p
\sim m_\pi$ from the ${\cal A}_{-1}^{n-2} f_{n,n-2}$ term. But these will cancel in
the $Q$ expansion of $p \, {\rm cot} \, \delta$.  Note that a calculation of the order
$Q_r^5$ graphs would be necessary to determine the order $Q^{3/2}$ terms.

\begin{figure}[!t]
  \centerline{\epsfxsize=18.0 cm \epsfbox{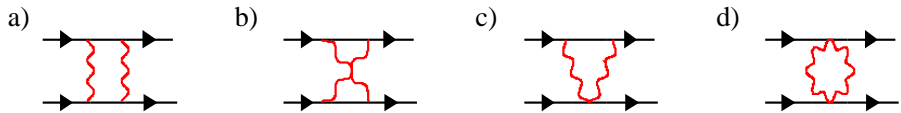}  }
{ \caption[1]{Examples of one-loop graphs which have soft pion contributions. 
Graphs a)-d) also have a radiation pion contribution, while in addition
graph a) has a potential pion contribution. } \label{fig_soft} }
\end{figure}
Finally, we briefly discuss soft contributions. These first arise in box-type diagrams
with two pions, such as those shown in Fig.\ \ref{fig_soft}. Dressing these graphs
on the outside with $C_0$ bubbles gives further diagrams of the same order. 
Unlike radiation graphs, which are dominated by loop momenta $|\vec{q}\,| \sim
m_\pi$, the box-like diagrams receive nonvanishing contributions from $|\vec{q}\,|
\sim Q_r$. It is this latter type of contribution which is called soft. In soft loops the
measure $d^4q \sim Q_r^4$ instead of $Q_r^8/M^4$.  For nucleon propagators in
soft loops, the loop energy is always greater than the nucleon's kinetic energy, so
the propagator is static (like heavy quark effective theory propagators, see
\cite{neubert} or \cite{gries}).  These propagators scale as $1/Q_r$.  Pion
propagators scale as $1/Q_r^2$ and pion-nucleon vertices give a factor of $Q_r$. 
Power counting the graphs in Fig.\ \ref{fig_soft} gives $\sim Q_r^2$ for the soft
contribution and $\sim Q_r^4/M^2$ for the radiation contribution. The contribution
from each regime can be separated using the method of asymptotic expansions
\cite{beneke,smirnov,smirnov2} and explicit evaluation \cite{radpi} of the graphs
verifies the power counting appropriate for each type of contribution. At $p \sim
Q_r$, the soft contribution actually dominates all radiation pion graphs we have
considered so far.  However, for $p\sim m_\pi$ these soft graphs are order $Q^2$,
and therefore are not enhanced by the scaling down to $p\sim m_\pi$. The leading
order soft graphs are ${\rm N}^3{\rm LO}$ in the KSW power counting.  


\section*{NNLO Calculation of the $^1S_0$ Phase Shift\footnote{The work 
presented in this section was done in collaboration with Sean Fleming.}}

In this section, we present a partial NNLO calculation of the $^1S_0$ phase shift. 
A more detailed analysis will be given in a future publication.\cite{fms} The first
piece of the $Q_r^3$ radiation pion contribution in Eq.~(\ref{fans}) is order $Q$. 
The second term is order $Q^{3/2}$ and is not included in the NNLO calculation. 
The relativistic corrections are computed in Ref.\cite{crs} and shown to be
negligible.  This is because they are suppressed relative to the leading order
amplitude by $(Q/M)^2$ rather than $(Q/\Lambda)^2$ and so are smaller than other
NNLO corrections.  The calculation here includes order $Q$ contact interactions
and potential pion graphs.  The NNLO calculation is incomplete because of the
omission of the $Q_r^4$ radiation pion graphs, as discussed in the previous
section. The pieces of the NNLO amplitude which are included are expected to
scale as $Q/(M \Lambda^2)$ while the $Q_r^4$ radiation pion graphs are expected
to scale as $Q/(\Lambda_\chi^2 \Lambda)$ for $p \sim m_\pi$. Since $\Lambda <
\Lambda_\chi \approx M$, the contribution from $Q_r^4$ radiation pion graphs may
be smaller than what has been included.  

Since we are only interested in the $^1S_0$ channel, only $s=S$ operators are 
relevant and  this superscript will be omitted in the following discussion.
At NNLO, the following terms are added to the Lagrangian in Eq.~(\ref{Lag}): 
\begin{eqnarray}
&{\cal L}=& -{C_4\over 64} \left[
 ( N^T P_i N)^\dagger ( N^T P_i\:\twoder^{\,4} N) + h.c. +
2( N^T P_i \twoder^{\,2} N)^\dagger ( N^T P_i\:\twoder^{\,2} N) \right] \nn \\
&&+{E_4 \over 8} \omega {\rm Tr}(m^\xi )
 \left[ ( N^T P_i N)^\dagger ( N^T P_i\:\twoder^{\,2} N) + h.c. \right] \nn \\
&&-{D_4 \over 2} \omega^2 \Big\{ {\rm Tr}^2(m^\xi )+2{\rm Tr}[(m^\xi)^2] \Big\}
 ( N^T P_i N)^\dagger ( N^T P_i N)  \,,
\end{eqnarray}
where only terms relevant for the phase shift are included and isospin violation is
neglected.  All calculations presented in this section will be done in the PDS
renormalization scheme\,\cite{ksw1,ksw2} with spin and isospin traces done in four
dimensions.

There are six coefficients that appear at NNLO: $C_0$, which is present in the
leading order calculation, $C_2$ and $D_2$, which first appear at NLO, and $C_4,
E_4$, and $D_4$.  It is important that the coupling constants are expanded in $Q$:
\cite{ms0}
\begin{eqnarray}\label{pc}
C_0 &\rightarrow& C_0 + C_{0,0} + C_{0,1} \nn\\
C_2 &\rightarrow& C_2 + C_{2,-1} \nn \\
D_2 &\rightarrow& D_2 + D_{2,-1} \,.
\end{eqnarray}
The first piece of $C_0$ is treated nonperturbatively (i.e. $C_0 \sim Q^{-1}$), while
$C_{0,0} \sim Q^0, C_{0,1} \sim Q$.  Solving the renormalization group equations
(RGE) for the couplings perturbatively ensures that the amplitude is $\mu$
independent order by order in the expansion. Therefore, theoretical expressions for
physical quantities, such as the scattering length or the location of the pole, are
always $\mu$ independent. Physically, the $Q$ expansion of the couplings is a
consequence of the fact that higher order loop graphs with pions can renormalize
the short distance operators at different orders in $Q$, and therefore different 
values of the couplings will be obtained at different orders in the expansion.  

When a coupling is expanded in $Q$ and its RGE solved perturbatively, a new
constant of integration is obtained for each term in the expansion. For example, the
RGE for $C_0$ is:
\begin{eqnarray} \label{C0RGE}
\mu {\partial \over \partial \mu} C_0 &=& {M \mu \over 4 \pi} 
\left( C_0 + {g_A^2 \over 2 f^2} \right)^2 \,. 
\end{eqnarray}
After the perturbative expansion of $C_0$ this becomes
\begin{eqnarray}
\mu {\partial \over \partial \mu} C_0 = {M \mu \over 4 \pi} C_0^2 \,, \qquad 
\mu {\partial \over \partial \mu} C_{0,0} = 2 {M \mu \over 4 \pi} C_0
\left( C_{0,0} + {g_A^2 \over 2 f^2} \right), \nn \\
\mu {\partial \over \partial \mu} C_{0,1} = {M \mu \over 4 \pi} \left[ 2 \,C_0 \,C_{0,1}
+\left( C_{0,0} + {g_A^2 \over 2 f^2} \right)^2 \: \right], \,\,\,\,\,\, \nn 
\end{eqnarray}
with solutions
\begin{eqnarray} \label{C0soln}
C_0 = {4 \pi \over M}{1 \over -\mu + \gamma} \,, \qquad
   C_{0,0} = {M \kappa \over 4\pi} \,C_0^2 - {g_A^2 \over 2 f^2} \,, \nn \\
 C_{0,1} =  {1 \over C_0}\left( C_{0,0} + {g_A^2 \over 2 f^2} \right)^2  
  +{M \kappa' \over 4\pi}  \: C_0^2 \,.  
\end{eqnarray}
Three constants of integration appear: $\gamma,\kappa$ and $\kappa'$; one for
each term in the $Q$ expansion of $C_0$.  For consistency we must assign these
constants a $Q$ counting.  For instance, $C_{0,0}$ has a term in its solution
$\kappa \, C_0^2 \sim \kappa / \mu^2$.  Since $C_{0,0} \sim Q^0$, $\kappa$ must be
of order $Q^2$. This reflects the fact that $\kappa$ is intrinsically small. In the
theory without pions, $\kappa = \gamma - 1/a$. As we will see below, the values of
$\gamma, \kappa, \kappa'$ may be fixed by demanding that the amplitude has the
correct pole structure. In this case, $\kappa \approx r_0/(2 a^2) \sim Q^2$. The 
results in Eq.~(\ref{C0soln}) can also be obtained by solving Eq.~(\ref{C0RGE})
exactly and expanding the result (including the constant of integration) in $Q$.
Because of the perturbative expansion of the couplings in Eq.~(\ref{pc}) there are
ten constants of integration at NNLO. However, the NNLO amplitude will depend
only on six independent linear combinations of these constants.  There are two
further constraints on the number of free parameters: 1) at this order, $C_4$, $E_4$
and $D_4$ are determined entirely in terms of lower order couplings; 2) spurious
double and triple poles in the NLO and NNLO amplitudes must be cancelled in
order to obtain a good fit.  

The fact that $C_4$, $E_4$ and $D_4$ are determined in terms of lower order
couplings is a consequence of solving the RGE's and applying the KSW power
counting.  For instance\,\cite{ksw2}, the RGE for $C_4$ is: 
\begin{eqnarray} 
  \mu {\partial \over \partial \mu} C_4 &=& {M \mu
  \over 4 \pi} \left(2 \,C_0 \,C_4 + C_2^2 \right) , \nn 
\end{eqnarray} 
which has the solution
\begin{eqnarray} 
   C_4 = {C_2^2 \over C_0} + \rho {M \over 4 \pi} C_0^2 \,, \nn 
\end{eqnarray}
where $\rho$ is a constant of integration. In the theory without pions, $\rho$ is
proportional to the shape parameter, which is $\sim Q^0$ in the KSW power 
counting. It is also reasonable to consider $\rho \sim Q^0$ in the theory with pions,
since $\rho$ is a constant of integration in the RGE for the lowest order term in the
$Q$ expansion of $C_4$.  Therefore, $C_2^2/C_0 \sim Q^{-3}$, while $\rho C_0^2
M/(4 \pi) \sim Q^{-2}$. The second term is subleading in the $Q$ expansion, and
should be omitted at NNLO, so $C_4=C_2^2/C_0$.  Similar relations for $E_4,
D_4$ hold at NNLO: 
\begin{eqnarray} 
  E_4 = {2 C_2 D_2 \over C_0} \,, \qquad\quad    D_4 = {D_2^2 \over C_0} \,. \nn 
\end{eqnarray} 
It is interesting to note that these relations arise even though there are 
$\ln(\mu^2)$ terms in the amplitude which contribute to the beta functions for
$E_4$ and $D_4$.

Because of the nonperturbative treatment of $C_0$, spurious poles arise at higher
orders in the expansion. The leading order amplitude ${\cal A}_{-1}$ has a simple
pole at $p= i \gamma$.  The NLO calculation is proportional to ${\cal A}_{-1}^2$,
and therefore has a double pole, while the NNLO amplitude has terms proportional
to ${\cal A}_{-1}^2$ and ${\cal A}_{-1}^3$. To obtain a good fit at low momentum,
parameters need to be fixed so that the amplitude has only a simple pole at each
order in the expansion. This requires that ${\cal A}_{-1}$ have its pole in the
correct location and that the residues of the spurious double and triple poles
vanish. This requirement leads to the following good fit conditions: \cite{ms0}
\begin{eqnarray}\label{goodfit} 
  \left. {1\over {\cal A}_{-1}}\right|_{p= p^*} \!\!\!\!\!\!\!\!\! = 0 \,, \qquad
  \left. {{\cal A}_0\over {\cal A}_{-1}^2}\right|_{p=p^*} \!\!\!\!\!\!\!\!\! = 0 \,,  \qquad
  \left. {{\cal A}_1\over {\cal A}_{-1}^2}\right|_{p=p^*} \!\!\!\!\!\!\!\!\!= 0 \,, \qquad
\end{eqnarray} 
where $p^*$ is the location of the physical pole.  The second condition first
appears at NLO, the third at NNLO.  The residue of the triple pole in ${\cal A}_1$ is
cancelled by the second equation in Eq.~(\ref{goodfit}). The first equation results in
$\gamma = -i p^*$, while the other equations give constraints which eliminate two
of the remaining parameters.  Eq.~(\ref{goodfit}) will also apply in the $^3S_1$ 
channel.

\begin{figure}[!t]
  \centerline{\epsfysize=9.0 cm \epsfbox{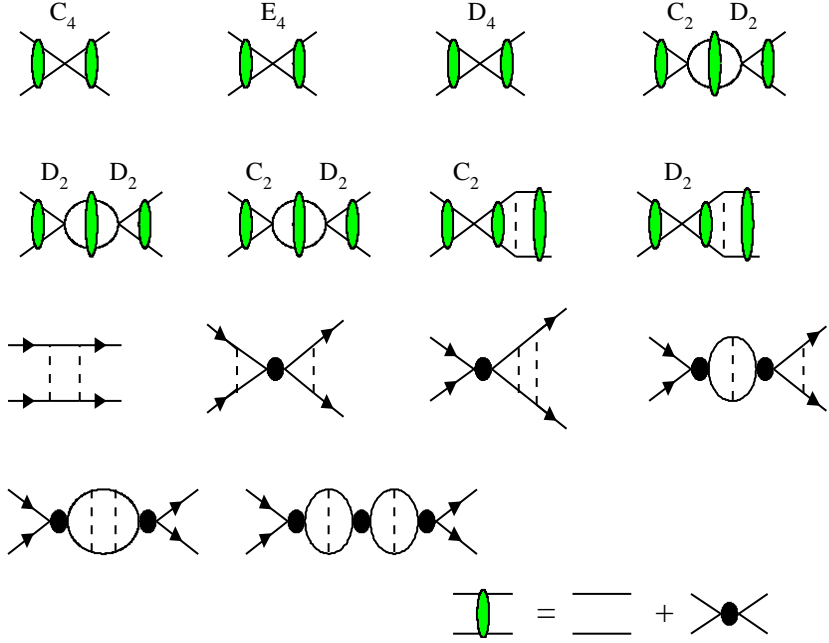}  }
{ \caption[1]{Order $Q$ contact interaction and potential pion graphs for the 
$^1S_0$ channel.  The shaded circle denotes a $C_0$ bubble sum.  At this order
the first six graphs cancel each other as explained in the text.} \label{figQ} }
\end{figure}
The graphs evaluated in the NNLO calculation are shown in Fig.\ 6. The final result
is surprisingly simple.  The amplitude up to NNLO is:
\begin{eqnarray} \label{Amp}
{\cal A}_{-1} &=& -{4\pi\over M}{1\over \gamma + ip}\,,  \nn \\[5pt]
{ {\cal A}_0  } &=& -  {\cal A}_{-1}^2 (\zeta_1 \, p^2 + \zeta_2 \, m_\pi^2)  \\*
   &&+ {g_A^2 \over 2 f^2} {\cal A}_{-1}^2 \Big({M m_\pi \over 4 \pi}\Big)^2 \bigg[ 
  { (\gamma^2 - p^2) \over 4 p^2}{\rm ln}\Big(1 + {4 p^2 \over m_\pi^2}\Big) - \, 
  {\gamma \over p} {\rm tan}^{-1}\left({2  p \over m_\pi} \right) \bigg] \nn\,, \\[5pt]
{ {\cal A}_{1} } &=& {{\cal A}_0^2 \over {\cal A}_{-1}} -  {\cal A}_{-1}^2
  \Big(\zeta_3 \, m_\pi^2 +\zeta_4 \,p^2 + \zeta_5 {p^4 \over m_\pi^2} \Big) 
  + { {\cal A}_0 }{M g_A^2 \over 8 \pi f^2}\,  {m_\pi^2\over p} 
  \bigg[ { \gamma \over 2 p}\: {\rm ln}\Big(1 + {4 p^2 \over m_\pi^2}\Big)  \nn \\*
&& - {\rm tan}^{-1}\Big({2 p \over m_\pi}\Big) \bigg]
  +{M {\cal A}_{-1}^2 \over 4 \pi}  \Big({ M g_A^2 \over 8\pi f^2}\Big)^2 
   {m_\pi^4 \over 4 p^3}  \Bigg\{ 2(\gamma^2-p^2)\, {\rm Im}\, {\rm Li_2} \Big(
   {-m_\pi \over m_\pi-2 i p} \Big)  \nn \\*
&&  - 4 \gamma\, p\: {\rm Re}\, {\rm Li_2}\Big({-m_\pi \over m_\pi-2 i p}
   \Big) - { \gamma\, p\, \pi^2 \over 3} -(\gamma^2+p^2) \bigg[ {\rm Im} \,
  {\rm Li_2} \Big({m_\pi+2 i p \over -m_\pi+2 i p} \Big) \nn \\*
&& + {\gamma \over 4 p} \ln^2\Big(1+\frac{4p^2}{m_\pi^2}\Big) - \tan^{-1}
   \Big(\frac{2p}{m_\pi} \Big) \ln\Big(1+\frac{4p^2}{m_\pi^2}\Big) \bigg] \Bigg\} \nn \,.
\end{eqnarray}
Using this amplitude it is easy to verify that the S-matrix is unitary to the order
we are working.
The six linearly independent constants appearing in the amplitude are
$\gamma,\zeta_1, \zeta_2,\zeta_3,\zeta_4,\zeta_5$. By definition $\zeta_1 -\zeta_5$
are dimensionless. They are given in terms of coupling constants in the Appendix.

For the $^1S_0$ channel, the location of the pole is determined by solving 
\begin{eqnarray}\label{pole}
    -{1 \over a} + {r_0 \over 2}(p^*)^2 - i p^* = 0 \,. 
\end{eqnarray}
Adding the shape parameter correction to Eq.~(\ref{pole}) changes the location of
the pole by less than $0.01\%$, so this approximation for the physical pole location
is sufficiently accurate.  Eq.~(\ref{pole}) has a solution for $p^* = - i \, 7.877 \,{\rm
MeV}$, which fixes the single LO parameter, $\gamma = - 7.877 \,{\rm
MeV}$.  The NLO good fit condition relates the constants $\zeta_1$ and $\zeta_2$,
\begin{eqnarray}\label{NLOfit}
  -\zeta_1 \gamma^2 &+& \zeta_2 m_\pi^2 +{M m_\pi^2 \over 4 \pi}{g_A^2 M \over 
  8 \pi f^2}  \left[ {1\over 2} {\rm log}\left(1-{4 \gamma^2 \over m_\pi^2}\right)-
  {\rm tanh}^{-1}  \left({2 \gamma \over m_\pi}\right)\right] = 0 \nn \\
&\Rightarrow& \zeta_2 = {\gamma^2 \over m_\pi^2}\, \zeta_1 + {M \over 4\pi}
   {g_A^2 M \over 8 \pi f^2}\left[ -{2 \gamma \over m_\pi} - {2 \gamma^2 \over 
   m_\pi^2 }  + {\cal O}\Big( {\gamma^3 \over m_\pi^3}\Big) \right] \,.
\end{eqnarray}
We can use this equation to eliminate $\zeta_2$ in favor of $\zeta_1$, leaving one
new parameter in the fit at NLO.  This good fit condition gives non-trivial 
$m_\pi$ dependence to the perturbative contributions to $C_0$ (such as $\kappa$) 
as emphasized in Refs.~\cite{Kaplan,NNLO,steele}.  $\zeta_1$ is fixed by doing a 
weighted least squares fit to low momentum data.  Note that $\zeta_1$ appears in
the good fit condition multiplied by $\gamma^2$. Therefore, the value of $\zeta_2$
is insensitive to the value of $\zeta_1$ obtained from the fitting procedure.  To a
good degree of accuracy we can ignore $\zeta_1$ in Eq.~(\ref{NLOfit}), and then we
find $\zeta_2 \approx 0.03$. $\zeta_2$ is small because it is proportional to
$\gamma/m_\pi$.  

At NNLO, $\zeta_5 = 0$ once we impose $C_4 = C_2^2/C_0$.  The condition in
Eq.~(\ref{NLOfit}) must still be satisfied. The NNLO good fit condition involves 
$\zeta_3$ and $\zeta_4$,
\begin{eqnarray} \label{NNLOfit}
 \zeta_3 &=& \frac{\gamma^2}{m_\pi^2}\,  \zeta_4 +  \Big(  {M g_A^2 \over 8\pi f^2}
   \Big)^2 {M\over 4\pi}\ {m_\pi^2 \over \gamma} \bigg[ {\rm Re}\,{\rm Li} 
   \Big({-m_\pi \over m_\pi+2 \gamma} \Big) + \frac{\pi^2}{12} \bigg]  \\
  &=& \frac{\gamma^2}{m_\pi^2}\,  \zeta_4 + \Big(  {M g_A^2
   \over 8\pi f^2}\Big)^2 {Mm_\pi \over 4\pi}  \bigg[ 2\ln2 - (1+2\ln2) 
   \frac{\gamma}{m_\pi}  + {\cal O}\Big( {\gamma^2 \over m_\pi^2} \Big) \bigg]
   \nn \,.
\end{eqnarray} 
Since $\zeta_4$ is multiplied by $\gamma^2/m_\pi^2$, this condition basically fixes 
the value of $\zeta_3$.  At this order $\zeta_1$ may change from its value at NLO. 
We have chosen to fix $\zeta_1$ and $\zeta_4$ by performing a least square fit to
the lower momentum data.

\begin{figure}[!t]  
  \centerline{\epsfxsize=6.5truecm \epsfbox{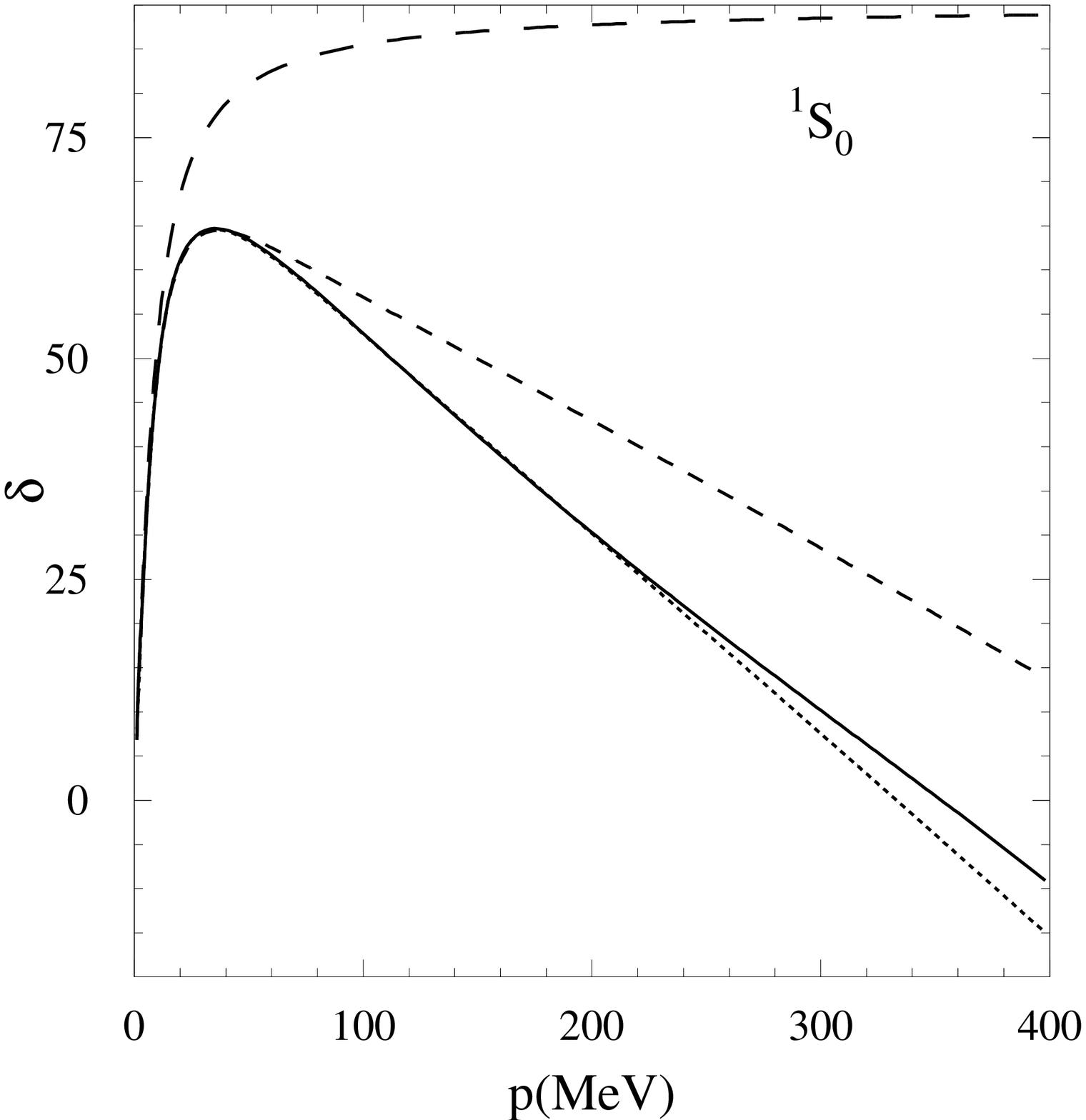}
        \epsfxsize=6.5truecm \epsfbox{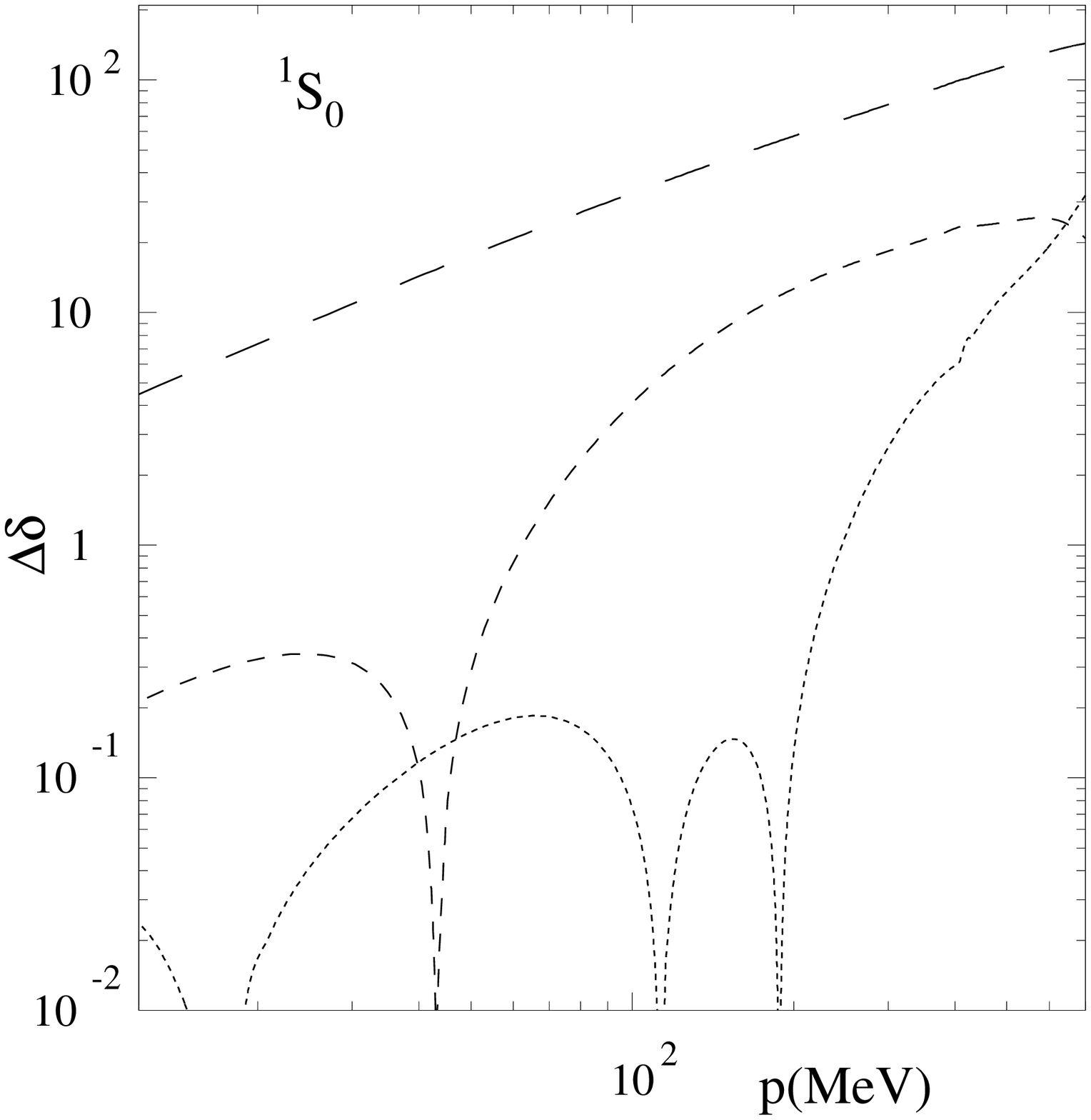} } 
\caption[1]{Fit to the $^1S_0$ phase shift $\delta$ emphasizing the low momentum
region.  The solid line is the Nijmegen fit \cite{Nij} to the data (for $p>400\,
{\rm MeV}$ values from the VPI \cite{VPI} phase shift analysis were used.). The 
long dashed, short dashed, and dotted lines are the LO, NLO, and NNLO results 
respectively. $\Delta \delta$ is the difference between these results and the solid 
line.} \label{fig_fits} 
\end{figure}
The $^1S_0$ phase shift is shown in Fig.~\ref{fig_fits}. The solid line is the result of
the Nijmegen phase shift analysis\,\cite{Nij}.  The $^1S_0$ phase shift has an
expansion in powers of $Q$, and we have plotted the LO, NLO and NNLO results.
The LO phase shift at $p \sim m_\pi$ is off by $49 \%$.  At NLO, the error is
$14 \%$. At NNLO, the error in the $^1S_0$ channel is less than $2\%$ at 
$p \sim m_\pi$, and the NNLO result gives improved agreement with the data 
even at $p \sim 400 \,{\rm MeV}$.  

Using $M=939\,{\rm MeV}$ and $m_\pi=137\,{\rm MeV}$, the parameters for 
our fit in the $^1S_0$ channel are:
\begin{eqnarray}
  {\rm NLO}: && \qquad \zeta_1 = 0.2163; \qquad \zeta_2 = 0.0318 ; \nn \\*
  {\rm NNLO}: && \qquad \zeta_1 = 0.0777 ; \qquad \zeta_2 = 0.0313 ;
   \qquad \zeta_3 = 0.1831 ; \qquad \zeta_4 = 0.2447 ; \nn  
\end{eqnarray}
Note that $\zeta_3\sim Q$ is larger than $\zeta_2\sim Q^0$ because from
Eqs.~(\ref{NLOfit}) and (\ref{NNLOfit}), $\zeta_3/\zeta_2 \sim m_\pi^2/
(\gamma\Lambda_{NN})$.  The parameter $\zeta_2$ is stable because it is fixed by
the NLO good fit condition.  On the other hand, $\zeta_1$ changes by a factor of
2.7 going from NLO to NNLO.  One expects the value of coupling constants to
change at each order in the expansion, but a factor of three difference is somewhat
surprising. It is also disturbing that $\zeta_4$ is greater than $\zeta_1$, since, on
the basis of the RGE, it is expected that $\zeta_4 < \zeta_1$.  At NNLO the RGE for
$C_2$ is: \cite{ms1}
\begin{eqnarray}
    \mu {\partial \over \partial \mu} C_2 &=& 2 {M \mu \over 4 \pi} \left( C_0 + 
   {g_A^2 \over 2 f^2}\right) C_2  \,. \nn 
\end{eqnarray}
Expanding $C_2$ perturbatively results in two equations:
\begin{eqnarray} \label{RGE2}
  \mu {\partial \over \partial \mu} C_2 &=& 2 {M \mu \over 4 \pi} \,C_0 \,C_2 \,,  \\
  \mu {\partial \over \partial \mu} C_{2,-1} &=& 2 {M \mu \over 4 \pi} 
  \left[ \left( C_{0,0} + {g_A^2 \over 2 f^2} \right) C_2 + C_0 \,C_{2,-1} \right] \,, \nn 
\end{eqnarray}
with solutions 
\begin{eqnarray} \label{C2S}
  C_2 = \zeta_1 \, C_0^2 \,, \qquad\quad C_{2,-1} = 2{C_2 \over C_0} \left( C_{0,0} 
  + {g_A^2 \over 2 f^2} \right)+ \zeta_4 \, C_0^2 \,. 
\end{eqnarray}
The second term in $C_{2,-1}$ has exactly the same form as the leading $C_2$. 
$C_{2,-1}$ is supposed to be a perturbative correction to $C_2$, so $\zeta_4\sim Q
< \zeta_1\sim Q^0$, and one does not expect this part of $C_{2,-1}$ to be
significantly larger than the leading $C_2$.

Some insight into this puzzle can be obtained by comparing the NLO and NNLO 
expressions for $r_0$.  Using Eq.~(\ref{Amp}) we have
\begin{eqnarray}
r_0^{NLO} &=& {8 \pi \over M} \zeta_1 + {2 \over \Lambda_{NN}} 
   \left(1 - {8 \gamma \over 3 m_\pi} + {2 \gamma^2 \over m_\pi^2} \right) \nn  \\
 &=& {8 \pi \over M} ( \zeta_1 + 0.2952 ) \,, \\ 
r_0^{NNLO} &=& {8 \pi \over M} \left[\zeta_4 +  \zeta_1 \left( 1 - {2 m_\pi-2\gamma 
   \over \Lambda_{NN}} \right) + { \zeta_2  \over \Lambda_{NN}}
   \left({8 m_\pi \over 3} - 4 \gamma \right) + \ldots \right] \nn \\ 
&=& {8 \pi \over M} ( \zeta_4 + 0.01375 \,\zeta_1 + 1.35078 \,\zeta_2 +0.210398 )\,,
  \label{r0} 
\end{eqnarray}
where $1/\Lambda_{NN} = (M g_A^2)/(8 \pi f^2)$. Any reasonable fit for the phase
shifts will at least approximately reproduce the observed effective range.  In 
Eq.~(\ref{r0}), the piece of the NLO correction proportional to $\zeta_1$ is almost
exactly cancelled by the NNLO correction. This cancellation occurs because
$1-2(m_\pi-\gamma)/\Lambda_{NN} \simeq 0.01$ in the $^1S_0$ channel. This is
simply an unfortunate accident; in the $^3S_1$ channel, where $\gamma = 45.7
\,{\rm MeV}$ instead of $- 7.88\, {\rm MeV}$, the coefficient of $\zeta_1$ in 
Eq.~(\ref{r0}) is $\approx 0.4$.  Since $\zeta_2$ is small due to the NLO good
fit condition, this accidental cancellation forces $\zeta_4$ to make up the observed
effective range. Therefore, $\zeta_4$ is much larger than anticipated.

It is important to note that the coupling $C_2$ is not changing nearly as drastically
at each order. If one were to solve the theory exactly, one would find that $C_2$
had a term ${\hat r} \, C_0^2$, where ${\hat r}$ represents a short distance
contribution to the effective range. $\zeta_1$ and $\zeta_4$ can be thought of as
the first few terms in an expansion of ${\hat r}$. The theory should eventually
converge to the exact ${\hat r}$ but it need not reproduce ${\hat r}$ exactly at low
orders in perturbation theory.  It is reassuring that $\zeta_1^{NLO} = 0.22$ and
$\zeta_1^{NNLO} + \zeta_4^{NNLO}= 0.32$, indicating that the coupling constant
$C_2$ is not changing more than one would expect in a theory with an expansion
parameter $\sim 1/3$.  Note that $\zeta_1^{NNLO}$ must be small in order for
$\zeta_1^{NNLO}+\zeta_4^{NNLO}$ to not be much larger than $\zeta_1^{NLO}$. 
At NLO potential pions make up $\sim 60\%$ of $r_0$, with short distance physics
making up the remaining $\sim 40\%$.  At NNLO the situation does not change by
very much; potential pions give $\sim 40\%$, short distance physics $\sim 50\%$
and cross-terms make up the rest.

Kaplan and Steele\,\cite{Kaplan,sk} have proposed a fitting procedure in which
${\hat r}$ is not expanded in powers of $Q$. This amounts to imposing the
additional condition $\zeta_4 =0$, so there is no new parameter at NNLO.  Only the
linear combination $\zeta_1+\zeta_4$ appears in $C_2$.  However the amplitude
depends on $\zeta_1$ and $\zeta_4$ very differently because they appear at
different orders in the $Q$ expansion.  This is why we treat $\zeta_1$ and $\zeta_4$
as seperate parameters.  Where it not for the cancellation in $r_0^{NNLO}$ noted
above, then the difference between the two methods would be small, i.e., the size
of a ${\rm N}^3{\rm LO}$ correction. In fact, it is impossible to reproduce the
observed effective range if one demands $\zeta_4 = 0$. In the $^3S_1$ channel
their is a logarithmic divergence\,\cite{ms1} at order $Q$,  introducing a
$\ln(\mu/K)$ dependence into the coupling $C_{2,-1}$.  Since the constant $K$ is
undetermined, $\zeta_4$ cannot be set to zero, so there is a new parameter at
NNLO.  For $C_2^S$ this type of $\ln(\mu)$ dependence occurs at order $Q^2$
from soft pion graphs.\cite{radpi}   Kaplan and Steele have suggested that the
failure of their fitting procedure when applied to models with effective ranges close
to that seen in nature may indicate that $r_0$ is unnaturally large, and that the
power counting of the effective theory might need to be modified to take this scale
into account.  It seems more likely that the failure observed in Ref.~\cite{sk} may
just be the consequence of a numerical accident in the amplitude at NNLO as
shown in Eq.~(\ref{r0}). This cancellation does not occur in the $^3S_1$ channel at
NNLO, nor is it likely to persist at higher orders. For this reason, it seems
premature to conclude on the basis of the NNLO amplitude that the expansion is
failing due to a large $r_0$.

In Ref.~\cite{NNLO2}, Rupak and Gautam use a similar fitting procedure to the 
one discussed here.  Instead of finding $\zeta_1$ and $\zeta_4$ by fitting to the 
phase shift, they fix these constants by matching onto 
\begin{eqnarray}
   p\, \cot(\delta) = -\gamma + \frac{s_0}2 \, (p^2+\gamma^2) + \ldots \,,
\end{eqnarray}
which is similar to the effective range expansion except $p\cot(\delta)$ is expanded
about $p=i\gamma$.  At NLO $\zeta_1$ is fixed to give $s_0$.  At NNLO the same
value of $\zeta_1$ is used and $\zeta_4$ is again fixed to reproduce $s_0$. This
procedure was applied to data from a two-Yukawa toy model and the convergence
of the EFT looks similar to that in Fig.~\ref{fig_fits}.  In our approach we have not
demanded that the exact value of $r_0$ is reproduced since we know that there
will be corrections to $r_0$ from higher orders in the $m_\pi/\Lambda$ expansion.

Finally, we would like to comment on the prediction of higher order terms in the
effective range expansion
\begin{eqnarray}
  p\,{\rm cot}(\delta) = -{1\over a} + {r_0 \over 2}p^2 + v_2 p^4 + v_3 p^4+ v_4 p^4 
     \ldots \nn  \,.
\end{eqnarray}
Using the NLO expression for $p\,{\rm cot}(\delta)$, Cohen and Hansen\,\cite{ch}
obtained predictions for $v_2,v_3$ and $v_4$.  At NLO, the effective field theory
predictions for $v_2$, $v_3$, and $v_4$ disagree with the $v_i$ obtained from a fit
to the Nijmegen phase shift analysis.  The NNLO predictions for the shape
parameters are shown in the table below.  The prediction for $r_0$ is not better at
NNLO than at NLO.  The NNLO $v_i$ predictions depend on $\zeta_1$ and
$\zeta_2$.  We see that the NNLO correction substantially reduces the discrepancy
between the effective field theory prediction and the fit to the Nijmegen phase shift
analysis, but the discrepancy is still quite large. This gives some evidence that the
EFT expansion is converging on the true values of the $v_i$, albeit slowly. 
Effective field theory predictions for the shape parameters have been studied in toy
models where one is able to go to very high orders in the $Q$
expansion.\,\cite{kaplanpc} In the toy models, the effective field theory did
eventually reproduce the shape parameters, but the observed convergence is
rather slow.
\begin{table}[h!] 
\begin{center} \begin{tabular}{ccccccc} \label{LET}
 $^1S_0$\hspace{2cm}& & & $r_0$ & $v_2$ & $v_3$ & $v_4$ \\ \hline 
&Fit\cite{ch} && $2.73\,{\rm fm}$ & $-0.48\,{\rm fm^3}$ & $3.8\,{\rm fm^5}$ & 
	$-17\,{\rm fm^7}$  \\ 
&NLO && $2.65\,{\rm fm}$ & $-3.3\,{\rm fm^3}$ & $19\,{\rm fm^5}$ & 
	$-117\,{\rm fm^7}$  \\
&NNLO && $2.63\,{\rm fm}$ & $-1.2\,{\rm fm^3}$ & $2.9\,{\rm fm^5}$ & 
	$-0.7\,{\rm fm^7}$ 
\end{tabular} \end{center}  
\end{table}

What can we learn from Fig.~\ref{fig_fits} about the convergence of the KSW
expansion? It is pleasing to see a NNLO calculation reproducing the $^1S_0$
phase shift at $p \sim m_\pi$ with an accuracy of a few percent, and giving an
improved fit to the data even for larger momenta.  It is important to keep in mind
that the NNLO calculation of the phase shift is incomplete since there is a possible
contribution from order $Q_r^4$ radiation pion graphs.  Many other process
involving two nucleons can be examined at this order. Once enough processes are
calculated to NNLO, all parameters of the theory appearing at this order can be
extracted and it will be possible to make predictions with no free parameters. The
accuracy of these predictions will constitute a serious test of the KSW expansion
method.  

\section*{Acknowledgments}
We would like to thank  D. Kaplan, G. Rupak, and N. Shoresh, for useful discussion.
This work was supported in part by the Department of Energy under grant number 
DE-FG03-92-ER 40701. T.M. was also supported by a John A. McCone Fellowship.

\section*{Appendix }

Here we give the definitions of the constants appearing in the NNLO expression for
the amplitude in Eq.~(\ref{Amp}).  After solving the renormalization group
equations at this order one finds that all quantities in parentheses and curly
brackets are separately $\mu$ independent. The quantities in curly brackets
vanish at NNLO in the $Q$ expansion.  
\begin{eqnarray}
\gamma&=&\,\,{4\pi\over M C_0} + \mu \ ; \qquad\quad
\zeta_1 = \, \left( {C_2 \over C_0^2} \right);  \\
\zeta_2 &=& \left( {D_2 \over C_0^2 } -{g_A^2 \over 4 f^2} \Big( {M\over 4 \pi} 
   \Big)^2 \left[1+{\rm ln}\Big({\mu^2 \over m_\pi^2}\Big)\right] \right) + 
   \left({g_A^2/(2f^2) + C_{0,0} \over C_0^2 \:m_\pi^2 }\right) ; \nn \\
\zeta_3 &=& -{g_A^2 \over 2 f^2}  {M m_\pi \over 4 \pi}
    \left( {C_2 \over C_0^2}\right) +{1 \over m_\pi^2} \left( {C_{0,1} \over C_0^2} -
    {[g_A^2/(2f^2) +C_{0,0}]^2 \over C_0^3} \right)  \nn \\
 && + m_\pi^2 \left\{ {D_4 \over C_0^2} \!\!-\! {D_2^2\over C_0^3} \right\} \!\!+\! 
    \left({D_{2,-1} \over C_0^2} - {2 D_2 [g_A^2/(2f^2) +C_{0,0} ] \over C_0^3} + 
    {g_A^2 \over 2f^2}{\mu M \over 4\pi} {C_2 \over C_0^2}\right) \nn \\
 && + \left({\Delta D_2(\mu)\, m_\pi^2 \over C_0(\mu)^2} - 6 \, {g_A^2 m_\pi^2 
     \over (4 \pi f)^2}\,  {M \over 4\pi}\,\Big({1\over a^S} -{1\over a^T}\Big) 
     \bigg[{1\over 3} + {\rm ln}\Big( {\mu^2 \over m_\pi^2 }\Big) \bigg]  \right);\nn \\
\zeta_4 &=& \left({C_{2,-1} \over C_0^2} - {2\, C_2 \,[g_A^2/(2f^2) +C_{0,0} ]
     \over C_0^3} \right) + m_\pi^2\ \left\{{E_4 \over C_0^3}  - {2\, C_2 \, D_2 \over 
     C_0^2} \right\} ; \qquad\quad \nn \\
\zeta_5 &=& m_\pi^2\ \left\{ {C_4 \over C_0^2} - {C_2^2\over C_0^3} \right\}\,. \nn
\end{eqnarray}
$\zeta_1$ and $\zeta_4$ are short distance constants of integration of the RGE's in
Eq.~(\ref{RGE2}).  On the other hand, $\zeta_2$ and $\zeta_3$ are sums of
constants and variables that appear in the amplitude.  Note that the order $Q$
radiation pion contribution from order $Q_r^3$ graphs is constant and appears in
$\zeta_3$.  $\Delta D_2(\mu)$ is a correction to $D_2$ which cancels the $\ln(\mu)$
dependence from the $Q_r^3$ radiation pion graphs.

\end{document}